\documentclass[letterpaper, 10 pt, conference]{ieeeconf}  % Comment this line out if you need a4paper

\IEEEoverridecommandlockouts                              % This command is only needed if 
\usepackage{amsmath} % assumes amsmath package installed
\usepackage{amsfonts}
\usepackage{amssymb}
\usepackage{nccmath}
\usepackage{algorithm,algorithmic}

\usepackage{subcaption}
\usepackage{cite}

\usepackage{pgfplots}
\usepackage{tikz} 

\pgfplotsset{compat=newest}
\usetikzlibrary{decorations.pathmorphing} % noisy shapes
\usetikzlibrary{fit} % fitting shapes to coordinates
\usetikzlibrary{backgrounds} % drawing the background after the foreground
\usetikzlibrary{external}
\usetikzlibrary{pgfplots.groupplots}
\DeclareMathOperator\erf{erf}

\title{\LARGE \bf Chance-Constrained Trajectory Optimization for \\ Non-linear Systems with Unknown Stochastic Dynamics}

\author{Onur Celik$^{1}$, Hany Abdulsamad$^{2}$ and Jan Peters$^{2,3}$% <-this % stops a space
\thanks{*This work has received funding from the European Union’s Horizon 2020 research
	    and innovation program under grant agreement \# 640554.}% <-this % stops a space
\thanks{$^{1}$Onur Celik is with the Department of Computer Science,
        Universtit\"at T\"ubingen. {\tt\small mevluet-onur.celik@uni-tuebingen.de}}%
\thanks{$^{2}$Hany Abdulsamad and Jan Peters are with the Department of Computer Science, Intelligent Autonomous Systems,
		Technische Universtit\"at Darmstadt. {\tt\small \{abdulsamad, peters\}@ias.tu-darmstadt.de}}%
\thanks{$^{3}$Jan Peters is with the Max Planck Institute for Intelligent Systems.}
}	

\begin{document}

\maketitle
\thispagestyle{empty}
\pagestyle{empty}

\begin{abstract}
	Iterative trajectory optimization techniques for non-linear dynamical systems are among the most powerful and sample-efficient methods of model-based reinforcement learning and approximate optimal control. By leveraging time-variant local linear-quadratic approximations of system dynamics and reward, such methods can find both a target-optimal trajectory and time-variant optimal feedback controllers. However, the local linear-quadratic assumptions are a major source of optimization bias that leads to catastrophic greedy updates, raising the issue of proper regularization. Moreover, the approximate models' disregard for any physical state-action limits of the system causes further aggravation of the problem, as the optimization moves towards unreachable areas of the state-action space. In this paper, we address the issue of constrained systems in the scenario of online-fitted stochastic linear dynamics. We propose modeling state and action physical limits as probabilistic chance constraints linear in both state and action and introduce a new trajectory optimization technique that integrates these probabilistic constraints by optimizing a relaxed quadratic program. Our empirical evaluations show a significant improvement in learning robustness, which enables our approach to perform more effective updates and avoid premature convergence observed in state-of-the-art algorithms.
\end{abstract}

\vspace{0mm}

\section{Introduction}
Model-based reinforcement learning has played an important role in the latest surge of popular research interest in learning-control of autonomous systems \cite{deisenroth2013survey}. More specifically, trajectory-centric optimization techniques of non-linear dynamics have proven to be extremely sample efficient in comparison to model-free policy search approaches \cite{tassa2012synthesis, deisenroth2015gaussian, levine2016end}. 

%\begin{figure}[h!]
%	\centering
%	\includegraphics[width=0.7\columnwidth]{graphs/furuta/qube_2}
%	\caption{The Furuta Pendulum: Performing a swing-up on this highly non-linear dynamical system is a challenging control objective. Trajectory optimization methods like as DDP and iLQR are efficient techniques for solving such tasks. They suffer however in constrained stochastic environments. We propose to model state-action constraints as probabilistic chance constraints.}
%	\label{furuta_pendulum}
%\end{figure}

With the notable exception of \cite{deisenroth2011pilco}, model-based trajectory optimization techniques \cite{levine2014learning, abdulsamad2017state} are closely related to differential dynamic programming methods (DDP), initially presented in \cite{jacobson1970differential} and further generalized in \cite{todorov2005generalized}. DDP is a powerful approach for generating optimal trajectories with optimal time-variant feedback controllers. By relying on linear-quadratic approximations of the dynamics and reward around a nominal trajectory, DDP-based methods can leverage the local approximations to iteratively optimize both the trajectory and tracking feedback controllers in closed-form via dynamic programming \cite{anderson2007optimal}. This view of control has a computational advantage over direct optimization techniques such as collocation methods, which solve large optimization problems directly in the trajectory space and generally result only in open-loop control sequences \cite{von1992direct}.

However, despite the overwhelming success of DDP, it still suffers from multiple shortcomings. On the one hand, the greedy exploitation of poor local approximations of dynamics is a major problem that leads to premature convergence. This issue has been effectively addressed in recent research by proposing different schemes of regularization \cite{tassa2012synthesis, levine2014learning, abdulsamad2017state}. On the other hand, state and action constraints present a serious challenge, as they introduce hard non-linearities, that cannot be straightforwardly incorporated into the dynamic programming framework. The effect of constraints becomes more severe in settings where a global model is not available for automatic differentiation, hence requiring the linear approximation of the dynamics to be fitted online from samples.

We view these issues of DDP as interlocked. The inability of time-variant local linear models to consider state and action constraints results in updates that exploit unreachable parts of the state-actions space, leading to catastrophically poor linear-quadratic approximations in regions subject to hard non-linearities. Moreover, considering constraints becomes more challenging in scenarios with stochastic dynamics, in that the true state of the system is hidden and only available through sufficient statistics. Another crucial aspect in a stochastic setting is the infinite support of the noisy measurements, which results in the constraints being active over the whole state-action space.

To address these issues, we propose an augmented view of DDP that introduces the physical limits as probabilistic chance constraints linear in state and action. When considering time-variant linear-Gaussian approximations of the dynamic, we can relax the generally non-convex chance constraints by applying Boole's inequality. This relaxation allows us to formulate an additional quadratic program that forces the optimized nominal trajectory to stay in a feasible state-action region with high probability, all while considering the feedback gains optimized by DDP.

Several approaches to trajectory optimization for non-linear systems address the problem of constrained dynamics on different levels. In the domain of deterministic environments, Tassa et al. considered action box-constraints in \cite{tassa2014control}, while the authors in \cite{plancher2017constrained} introduce soft state-action limits via a Lagrange function augmentation. More sophisticated integration of constraints is presented in \cite{xie2017differential}, in which the authors formulate a quadratic program to determine the active set of constraints at every iteration. In a stochastic setting, the work by Van Den Berg et. al \cite{van2012efficient} introduces probabilistic constraints as direct penalty terms on the cost function.

Furthermore, probabilistic constraints are considered in the context of linear optimal control. In \cite{van2004stochastic} the authors optimally handle probabilistic constraints by ellipsoidal relaxation for finite-horizon open-loop scenarios, while in \cite{blackmore2009convex} a similar problem is tackled by applying Boole's inequality. In \cite{vitus2011closed} Vitus et al. propose an algorithm to extend the work in \cite{blackmore2009convex} and \cite{van2004stochastic} by considering closed-loop uncertainty and optimizing the risk allocation. Finally, in \cite{kurz2016progressive} the problem of infeasible initial solutions is addressed by progressively introducing the constraint into the objective.

We situate our contribution in the class of differential dynamic programming for stochastic non-linear systems subject to probabilistic constraints in state and action. We empirically show that our proposed approach can deal with highly non-linear constrained dynamic environments, leading to better overall performance and a robust learning process by virtue of improved online-fitted local approximations.

\section{Chance-Constrained Optimization}

Chance constraints arise naturally in different fields of optimization when considering stochastic systems. For an overview, we refer to \cite{prekopa2013stochastic}. Dealing with such probabilistic constraints proves to be challenging, as they are often non-convex and hard to evaluate without resolving to computationally expensive sampling techniques. These difficulties have motivated further research into tractable forms of chance constraints, which led to several convex approximations \cite{nemirovski2006convex}. This work will focus on using Boole's inequality for constraint relaxation. A detailed description in the context of trajectories will follow in Section~\ref{RelaxingCC}.

\subsection{Problem Formulation}
Consider the constrained optimal control problem with probabilistic state and action constraints and unknown stochastic time-discrete transition dynamics
\begin{align*}
	\underset{\boldsymbol{A}}{\text{max}}~~~ & J(\boldsymbol{s},\boldsymbol{A}),                                                                \\
	\text{s.t.}~~~                           & \boldsymbol{s}_{t+1} \sim \mathcal{P}(\boldsymbol{s}_{t+1}|\boldsymbol{s}_{t},\boldsymbol{a}_t), \\
	                                         & \Pr(\boldsymbol{s}_{0:T} \in \mathcal{S})\geq 1-\theta,                                          \\
	                                         & \Pr(\boldsymbol{a}_{0:T-1} \in \mathcal{A})\geq 1-\vartheta, 
\end{align*}
where $\mathcal{S} $ and $\mathcal{A} $ are the feasible state and action spaces respectively. The probability levels $\theta, \vartheta$ are hyperparameters that influence the risk behavior in terms of violating the constraints. The goal of this constrained optimization is to maximize the objective by finding the optimal action sequence $\boldsymbol{A} $. In general, we consider the expected cumulative reward for a trajectory of length $T$ in the quadratic form%
\begin{align}\label{quadrObj}
	J(\boldsymbol{s}, \boldsymbol{A}) = -\mathbb{E} \Big[ & \sum_{t= 0}^{T-1}(\boldsymbol{s}_t-\boldsymbol{s}_{g,t})^{\intercal}\boldsymbol{M}_t(\boldsymbol{s}_t-\boldsymbol{s}_{g,t}) + \boldsymbol{a}_t^{\intercal}\boldsymbol{D}_t\boldsymbol{a}_t \nonumber \\
	                                                      & + (\boldsymbol{s}_T-\boldsymbol{s}_{g,T})^{\intercal}\boldsymbol{M}_T(\boldsymbol{s}_T-\boldsymbol{s}_{g,T}) \Big],
\end{align}
where $\boldsymbol{M}$ and $\boldsymbol{D}$ are positive-definite weight matrices of appropriate dimensions and $\boldsymbol{s}_{g}$ is the target state. Note that a quadratic objective is not necessarily required, and non-quadratic objectives can be locally approximated by quadratic forms.

\subsection{Relaxation of Chance Constraints}\label{RelaxingCC}
Chance constraints can be conservatively relaxed by applying Boole's inequality \cite{boyd2004convex, vitus2011feedback, jha2016optimal}. For the purpose of brevity, only upper-bound state constraints are considered. However, the same relaxation procedure can be straightforwardly applied to obtain a lower-bound and to relax the action constraints. Generally, the state-linear joint chance constraint for a whole trajectory is formulated as
\vspace{-2mm}
\begin{align}\label{ChanceConstraintNotRelaxed}
	\Pr(\boldsymbol{s}_{0:T} \in \mathcal{S}) & = \Pr(\bigcap_{t=0}^{T} \boldsymbol{s}_t \in \mathcal{S}) \geq 1-\theta, \nonumber              \\
	                                          & = \Pr(\bigcap_{t=0}^{T} \boldsymbol{h}_t^{\intercal}\boldsymbol{s}_t  \leq b_t)  \geq 1-\theta.
\end{align}
where $\boldsymbol{h}_t$ and $b_t$ parameterize the half-plane defined by the constraints.
Consequently, the probability of a trajectory to be within a feasible set is constrained to be higher than a probability $1-\theta$. In the framework of stochastic programming, it is usually beneficial to reformulate Equation~\eqref{ChanceConstraintNotRelaxed} into separate inequalities over individual constraints \cite{prekopa2013stochastic}, which is achieved by transforming the intersection operator into a union operator according to rules of probability.
\vspace{-2mm}
\begin{align}\label{ProbSums}
	\Pr(\bigcap_{t=0}^{T} \boldsymbol{h}_t^{\intercal}\boldsymbol{s}_t  \leq b_t) & = 1-\Pr(\bigcup_{t=0}^{T} \boldsymbol{h}_t^{\intercal}\boldsymbol{s}_t  > b_t), \nonumber \\ 
	                                                                              & \geq  1 - \sum_{t=0}^{T}1-\Pr(\boldsymbol{h}_t^{\intercal}\boldsymbol{s}_t \leq b_t).
\end{align}
The sum in Inequality~(\ref{ProbSums}) can now be treated as a collection of single probabilities per time-step
\vspace{-2mm}
\begin{align}\label{SingleProbabilities}
	\sum_{t=0}^{T} 1-\Pr(\boldsymbol{h}_t^{\intercal}\boldsymbol{s}_t \leq b_t) & \leq \theta, \nonumber \\
	\Pr(\boldsymbol{h}_t^{\intercal}\boldsymbol{s}_t \leq b_t)                  & \geq 1-\theta_{t},
\end{align}
where $\sum_{t=0}^{T} \theta_{t} = \theta $. By assuming a Gaussian probability density, a common assumption in control applications, Equation~(\ref{SingleProbabilities}) is rewritten using the cumulative density function
\begin{align}\label{CCstate}
	\frac{1}{2}\left[1+ \erf \left(\frac{b_t - \boldsymbol{h}_t^{\intercal}\boldsymbol{\mu_{\boldsymbol{s}_t}}}{\sqrt{2\boldsymbol{h}_t^{\intercal}\boldsymbol{\Sigma}_{\boldsymbol{s}_t}\boldsymbol{h}_t}}\right)\right] \geq 1-\theta_{t},\nonumber \\[1mm]    
	b_t-\boldsymbol{h}_t^{\intercal}\boldsymbol{\mu_{\boldsymbol{s}_t}}-{\sqrt{2\boldsymbol{h}_t^{\intercal}\boldsymbol{\Sigma}_{\boldsymbol{s}_t}\boldsymbol{h}_t}}~ \erf^{-1}(1-2\theta_{t}) & \geq 0, 
\end{align}
where $\boldsymbol{\mu_{\boldsymbol{s}_t}}$ and $\boldsymbol{\Sigma}_{\boldsymbol{s}_t}$ are the state mean and covariance respectively. Moreover, due to properties of the error function, the inequality $\sum_{t=0}^{T} \theta_{t} \leq \theta < 0.5 $ is conservatively enforced by setting $\theta_{t} = \theta  / T$ and requiring $\theta < 0.5 $, as in \cite{jha2016optimal}.

\subsection{Iterative Linear Quadratic Gaussian Control (iLQG)}
We base our trajectory optimization technique on DDP/iLQG methods. This section provides a short overview on the principles of DDP \cite{jacobson1970differential} and iLQG \cite{tassa2012synthesis}.
For any arbitrary time-index reward function $R_t$, the trajectory optimization objective is the expected cumulative reward
\begin{align*}
	J(\boldsymbol{s},\boldsymbol{A}) = \mathbb{E}\left[\sum_{t=0}^{T-1} R_t(\boldsymbol{s}_t,\boldsymbol{a}_t) + R_T(\boldsymbol{s_T})\right] .
\end{align*}
DDP and iLQG leverage the principle of dynamic programming to simplify the optimization over a complete sequence of actions $\boldsymbol{a}_{0:T-1}$ to an optimization over single actions $\boldsymbol{a}_t $ for each time-step. For this purpose the time-indexed state-value function is introduced
\begin{equation*}
	V_t(\boldsymbol{s}) \! = \underset{\boldsymbol{a}_t}{\text{max}} \! \left[R_t(\boldsymbol{s}_t,\boldsymbol{a}_t) +\sum_{\boldsymbol{s}_{t+1}}V _{t+1}(\boldsymbol{s}_{t+1})\mathcal{P}(\boldsymbol{s}_{t+1}|\boldsymbol{s}_t,\boldsymbol{a}_t)\right] \! ,  
\end{equation*}
over which the dynamic programming backward recursion is performed.
By assuming linear transitions dynamics and a quadratic rewards along a nominal trajectory, optimal feedback controllers can be derived in closed-form. DDP and iLQG consider the perturbed state-action-value function $ Q_t(\delta \boldsymbol{s},\delta \boldsymbol{a}) = R_t(\boldsymbol{s}_t+\delta \boldsymbol{s},\boldsymbol{a}_t + \delta \boldsymbol{a}) - R_t(\boldsymbol{s}_t,\boldsymbol{a}_t) + V_{t+1} \left(\mathcal{P}(\boldsymbol{s}_t + \delta \boldsymbol{s}, \boldsymbol{a}_t + \delta \boldsymbol{a})\right) - V_{t+1}\left(\mathcal{P}(\boldsymbol{s}_t, \boldsymbol{a}_t)\right)$, resulting from a second order Taylor approximation
%\begin{align*}
%	Q_t(\delta \boldsymbol{s},\delta \boldsymbol{a}) = R_t(\boldsymbol{s}+\delta \boldsymbol{s},a + \delta \boldsymbol{a}) - R_t(\boldsymbol{s},\boldsymbol{a}) + \\
%	V' \left(\mathcal{P}(\boldsymbol{s} + \delta \boldsymbol{s}, \boldsymbol{a} + \delta \boldsymbol{a})\right) - V'\left(\mathcal{P}(\boldsymbol{s}, \boldsymbol{a})\right),
%\end{align*}
%where $ Q_t(\delta \boldsymbol{s},\delta \boldsymbol{a}) $ results from the second order Taylor approximation as
\begin{equation*}
	Q_t(\delta \boldsymbol{s},\delta \boldsymbol{a}) \approx \frac{1}{2}
	\begin{bmatrix}
		1                     \\
		\delta \boldsymbol{s} \\
		\delta \boldsymbol{a}
	\end{bmatrix}^{\intercal}
	\begin{bmatrix}
		0                    & \boldsymbol{Q}_{s,t}^{\intercal} & \boldsymbol{Q}_{a,t}^{\intercal} \\
		\boldsymbol{Q}_{s,t} & \boldsymbol{Q}_{ss,t}            & \boldsymbol{Q}_{sa,t}            \\
		\boldsymbol{Q}_{a,t} & \boldsymbol{Q}_{as,t}            & \boldsymbol{Q}_{aa,t}
	\end{bmatrix}
	\begin{bmatrix}
		1                     \\
		\delta \boldsymbol{s} \\
		\delta \boldsymbol{a}
	\end{bmatrix}.
\end{equation*}
The subscripts $s$ and $a$ stand for the first and second order approximations. The entries of $Q_t(\delta \boldsymbol{s},\delta \boldsymbol{a})$ are given as 
\begin{align*}
	\boldsymbol{Q}_{s,t}  & = \boldsymbol{R}_{s,t} + \mathcal{P}_{s,t}^{\intercal} \boldsymbol{V}_{\! s,t+1},                                                                  \\
	\boldsymbol{Q}_{a,t}  & = \boldsymbol{R}_{a,t} + \mathcal{P}_{a,t}^{\intercal} \boldsymbol{V}_{\! s,t+1} ,                                                                 \\
	\boldsymbol{Q}_{ss,t} & = \boldsymbol{R}_{ss,t} + \mathcal{P}_{s,t}^{\intercal} \boldsymbol{V}_{\! ss,t+1} \mathcal{P}_{s,t} +\boldsymbol{V}_{\! s,t+1}\mathcal{P}_{ss,t}, \\
	\boldsymbol{Q}_{aa,t} & = \boldsymbol{R}_{aa,t} + \mathcal{P}_{a,t}^{\intercal} \boldsymbol{V}_{\! ss,t+1} \mathcal{P}_{a,t} +\boldsymbol{V}_{\! s,t+1}\mathcal{P}_{aa,t}, \\
	\boldsymbol{Q}_{as,t} & = \boldsymbol{R}_{as,t} + \mathcal{P}_{a,t}^{\intercal} \boldsymbol{V}_{\! ss,t+1} \mathcal{P}_{s,t} +\boldsymbol{V}_{\! s,t+1}\mathcal{P}_{as,t}.
\end{align*}
The main difference of iLQG compared to DDP is in neglecting the second order derivatives of the dynamics in iLQG. Given these approximations the optimal feedback controller is given as $\delta \boldsymbol{a}^* = -\boldsymbol{Q}_{aa,t}^{-1}(\boldsymbol{Q}_a +\boldsymbol{Q}_{as,t}\delta \boldsymbol{s}) =  \boldsymbol{K}_t \delta \boldsymbol{s} + \boldsymbol{k}_t. $
Inserting $\delta \boldsymbol{a}^* $ into $ Q_t(\delta \boldsymbol{s},\delta \boldsymbol{a}) $ returns the update equations of the state-value function per time-step
\begin{align*}
	\Delta V_t               & = -\frac{1}{2}\boldsymbol{Q}_{a,t}\boldsymbol{Q}_{aa,t}^{-1}\boldsymbol{Q}_{a,t},                \\ 
	\boldsymbol{V}_{\! s,t}  & = \boldsymbol{Q}_{s,t} - \boldsymbol{Q}_{a,t}\boldsymbol{Q}_{aa,t}^{-1}\boldsymbol{Q}_{as,t},    \\
	\boldsymbol{V}_{\! ss,t} & = \boldsymbol{Q}_{ss,t} - \boldsymbol{Q}_{sa,t} \boldsymbol{Q}_{aa,t}^{-1}\boldsymbol{Q}_{as,t}.
\end{align*}
During the forward pass, new trajectories of the stochastic non-linear dynamics are sampled by propagating the actions through the real system
\begin{align}\label{fw_pass_iLQG}
	\boldsymbol{a}_t     & = \boldsymbol{a}_{r,t} + \boldsymbol{k}_t +\boldsymbol{K}_t (\boldsymbol{s}_t - \boldsymbol{s}_{r,t}),\nonumber             \\ 
	\boldsymbol{s}_{t+1} & \sim \mathcal{P}(\boldsymbol{s}_{t+1}|\boldsymbol{s}_{t},\boldsymbol{a}_t),  \quad \boldsymbol{s}_0 = \boldsymbol{s}_{r,0},
\end{align}
where $\boldsymbol{s}_{r,t}, \boldsymbol{a}_{r,t} $ denote the mean state and action at time $t$ from the last iteration and are also referred to as the nominal or reference trajectory, here denoted by the subscript $r$.

Special care has to be taken during the backward pass of DDP and iLQG to ensure that $\boldsymbol{Q}_{aa,t} $ is negative-definite, which has inspired different regularization schemes. In DDP, this regularization is commonly applied to $\boldsymbol{Q}_{aa,t}$ as $\boldsymbol{\tilde{Q}}_{aa,t} =  \boldsymbol{Q}_{aa,t} - \mu \boldsymbol{I}$, with $\mu \geq 0$. 
However, other regularizations directly affecting the value function have been shown to be more effective \cite{tassa2012synthesis}, and will be used throughout this work.

\subsection{Augmented Linearized Closed-Loop System}\label{sec:closed-loop}
To formulate the chance-constrained optimization problem, we first introduce the notation and system description of the online-fitted time-variant linear system. Following \cite{kurz2016progressive}, our approach optimizes the feedforward terms of the control, while satisfying the constraints for the linearized dynamics and maintains the feedback gains computed during the backward pass of DDP/iLQG.

Given $N$ trajectories from the non-linear system as described in Equation~\eqref{fw_pass_iLQG}, we fit linear-Gaussian models to the sampled data via regularized linear regression. Consequently we obtain the transition and control matrices $\boldsymbol{A}_t, \boldsymbol{B}_t$, as well as the bias vector $\boldsymbol{c}_t $ for each time-step. The resulting time-variant linear dynamics $\boldsymbol{s}_{t+1}  = \boldsymbol{A}_t \boldsymbol{s}_t + \boldsymbol{B}_t \boldsymbol{a}_t + \boldsymbol{c}_t + \boldsymbol{w}_t$, with $\boldsymbol{w}_t \sim \mathcal{N}(\boldsymbol{0}, \Sigma_t)$, and the controller $\boldsymbol{a}_t = \boldsymbol{K}_t (\boldsymbol{s}_t -\boldsymbol{s}_{r,t}) + \boldsymbol{k}_t +\boldsymbol{a}_{r,t} $ are used to formulate the closed-loop linear system $\boldsymbol{s}_{t+1}  = \boldsymbol{\hat{A}}_t \boldsymbol{s}_t + \boldsymbol{B}_t  \boldsymbol{k}_t + \boldsymbol{d}_t +\boldsymbol{w}_t$, where $\boldsymbol{\hat{A}}_t = \boldsymbol{A}_t+\boldsymbol{B}_t\boldsymbol{K}_t $ and $\boldsymbol{d}_t = \boldsymbol{c}_t - \boldsymbol{B}_t \boldsymbol{K}_t \boldsymbol{s}_{r,t} +\boldsymbol{B}_t \boldsymbol{a}_{r,t} $.

%\begin{align*}
%	\boldsymbol{s}_{t+1} & = \boldsymbol{A}_t \boldsymbol{s}_t + \boldsymbol{B}_t \boldsymbol{a}_t + \boldsymbol{w}_t +\boldsymbol{c}_t, \\
%	\boldsymbol{a}_t     & = \boldsymbol{K}_t (\boldsymbol{s}_t -\boldsymbol{s}_{r,t}) + \boldsymbol{k}_t +\boldsymbol{a}_{r,t}.
%\end{align*}

%Thus the perturbed linear closed-loop system can be formulated as  
%\begin{align*}
%	\boldsymbol{s}_{t+1} & = \hat{\boldsymbol {A}}_t \boldsymbol{s}_t + \boldsymbol{B}_t  \boldsymbol{k}_t + \boldsymbol{w}_t + \boldsymbol{d}_t,
%\end{align*}
%where $\boldsymbol{\hat{A}} = \boldsymbol{A+\boldsymbol{BK}} $ and $\boldsymbol{d}_t = \boldsymbol{c}_t - \boldsymbol{B}_t \boldsymbol{K}_t \boldsymbol{s}_{r,t} +\boldsymbol{B}_t \boldsymbol{a}_{r,t} $.

To represent the closed-loop system over an entire trajectory we use the augmented notation
\begin{align*}
	\boldsymbol{\tilde{s}} & \! = \!
	\begin{bmatrix}
		\boldsymbol{s}_0 \\
		\boldsymbol{s}_1 \\
		\vdots           \\
		\boldsymbol{s}_T \!
	\end{bmatrix}\! \!, \!
	\tilde{{\boldsymbol{k}}} \! = \!
	\begin{bmatrix}
		\boldsymbol{k}_0 \\
		\boldsymbol{k}_1 \\
		\vdots           \\
		\boldsymbol{k}_{T-1} \!
	\end{bmatrix}\! \!, \!
	\boldsymbol{\tilde{w}} \! = \!
	\begin{bmatrix}
		\boldsymbol{w}_0 \\
		\boldsymbol{w}_1 \\
		\vdots           \\
		\boldsymbol{w}_{T-1} \!
	\end{bmatrix}\! \!, \!
	\boldsymbol{\tilde{A}} =
	\begin{bmatrix}
		\boldsymbol{I}                                           \\
		\boldsymbol{\hat{A}}_0                                   \\
		\vdots                                                   \\
		\boldsymbol{\hat{A}}_{T-1} \cdots \boldsymbol{\hat{A}}_0 \\
	\end{bmatrix}\! \!, \!
	\\
	\boldsymbol{\tilde{B}} & \! = \!
	\begin{bmatrix}
		\boldsymbol{0}                                                            & \boldsymbol{0}                                                            & \dots  & \boldsymbol{0}       \\
		\boldsymbol{B}_0                                                          & \boldsymbol{0}                                                            & \dots  & \boldsymbol{0}       \\
		\boldsymbol{\hat{A}}_1 \boldsymbol{B}_0                                   & \boldsymbol{B}_1                                                          & \dots  & \boldsymbol{0}       \\
		\vdots                                                                    & \vdots                                                                    & \ddots & \vdots               \\
		\boldsymbol{\hat{A}}_{T-1} \cdots \boldsymbol{\hat{A}}_1 \boldsymbol{B}_0 & \boldsymbol{\hat{A}}_{T-1} \cdots \boldsymbol{\hat{A}}_2 \boldsymbol{B}_1 & \dots  & \boldsymbol{B}_{T-1}
	\end{bmatrix}\! \!, \!
	\\
	\boldsymbol{\tilde{d}} & \! = \!
	\begin{bmatrix}
		\boldsymbol{d}_0 \\
		\boldsymbol{d}_1 \\
		\vdots           \\
		\boldsymbol{d}_{T-1} \!
	\end{bmatrix}\! \!, \!
	\boldsymbol{\tilde{G}} \! = \!
	\begin{bmatrix}
		\boldsymbol{0}                                             & \boldsymbol{0}                                             & \dots  & \boldsymbol{0} \\
		\boldsymbol{I}                                             & \boldsymbol{0}                                             & \dots  & \boldsymbol{0} \\
		\boldsymbol{\hat{A}}_1                                     & \boldsymbol{I}                                             & \dots  & \boldsymbol{0} \\
		\vdots                                                     & \vdots                                                     & \ddots & \vdots         \\
		\boldsymbol{\hat{A}}_{T-1}  \cdots  \boldsymbol{\hat{A}}_1 & \boldsymbol{\hat{A}}_{T-1}  \cdots  \boldsymbol{\hat{A}}_2 & \cdots & \boldsymbol{I}
	\end{bmatrix}\! \!, \!
\end{align*}

The augmented weighting matrices for the quadratic objective take the form 
\begin{align*}
	\boldsymbol{\tilde{M}}     & = \text{diag}(\boldsymbol{M}_0, \dots, \boldsymbol{M}_{T}),~ \boldsymbol{\tilde{D}} = \text{diag}(\boldsymbol{D}_0, \dots, \boldsymbol{D}_{T-1}), \\
	\boldsymbol{\tilde{M}}_{C} & = \text{diag}(\boldsymbol{M}_0 +\boldsymbol{K}_{0}^{\intercal}\boldsymbol{D}_0\boldsymbol{K}_0, \dots,                                            \\
	                           & \qquad \quad ~ \boldsymbol{M}_{T-1} + \boldsymbol{K}_{T-1}^{\intercal} \boldsymbol{D}_{T-1}\boldsymbol{K}_{T-1},  \boldsymbol{M}_T),              \\
	\boldsymbol{\tilde{K}}     & = \text{diag}(\boldsymbol{K_0},\dots, \boldsymbol{K}_{T-1}),
\end{align*}
and the closed-loop linearized stochastic dynamics is written in terms of the augmented notation as
\begin{equation}\label{LinDynamcis}
	\boldsymbol{\tilde{s}} = \boldsymbol{\tilde{A}}\boldsymbol{s}_0 +  \boldsymbol{\tilde{B}} \boldsymbol{\tilde{k}} +  \boldsymbol{\tilde{G}} \boldsymbol{\tilde{w}} +  \boldsymbol{\tilde{G}} \boldsymbol{\tilde{d}},
\end{equation}

% M_mod_tilde als matrix:
%	\begin{align*}
%		 \boldsymbol{\tilde{M}}_{mod}&=
%		\begin{bmatrix}
%		\boldsymbol{M}_0 +\boldsymbol{K}_{0}^{\intercal}\boldsymbol{D}_0\boldsymbol{K}_0  			& \dots			&\boldsymbol{0}\\
%		\vdots																				&\ddots			&\vdots\\	
%		\boldsymbol{0}																		& \boldsymbol{M}_{T-1} + \boldsymbol{K}_{T-1}^{\intercal} \boldsymbol{D}_{T-1}\boldsymbol{K}_{T-1}  															& \boldsymbol{0}\\
%		\boldsymbol{0}																		& \boldsymbol{0}& \boldsymbol{M}_T
%		\end{bmatrix},
%	\end{align*}

% M_tilde as matrix:
%	\begin{align*}
%			 \boldsymbol{\tilde{M}} &=
%			\begin{bmatrix}
%			\boldsymbol{M}_0 				&\dots 			&\boldsymbol{0}\\
%			\boldsymbol{0}					& \dots 		&\boldsymbol{0}\\
%			\vdots				    		&\ddots 		&\vdots \\
%			\boldsymbol{0}					& \dots			&\boldsymbol{M}_{T}
%			\end{bmatrix},
%	\end{align*}

% D and K as matrix:
%	\begin{align*}
%			 \boldsymbol{\tilde{D}} &=
%			\begin{bmatrix}
%			\boldsymbol{D}_0 			& \boldsymbol{0}										&\dots 			&\boldsymbol{0}\\
%			\boldsymbol{0}			& \boldsymbol{D}_1									& \dots 		&\boldsymbol{0}\\
%			\vdots				& \vdots										&\ddots 		&\vdots \\
%			\boldsymbol{0}			& \dots											& \dots			&\boldsymbol{D}_{T-1}
%			\end{bmatrix}, ~~~  \boldsymbol{\tilde{K}}=
%			\begin{bmatrix}
%			\boldsymbol{K}_0 			&\boldsymbol{0}										&\dots 			&\boldsymbol{0}\\
%			\boldsymbol{0}			&\boldsymbol{K}_1 									&\dots			&\boldsymbol{0}\\
%			\vdots				&\vdots											&\ddots			&\vdots \\
%			\boldsymbol{0}			& \boldsymbol{0}										&\dots			&\boldsymbol{K}_{T-1}
%			\end{bmatrix}.
%	\end{align*}

which in turn can be decomposed to the mean and covariance  of a Gaussian state density
\begin{align*}
	\boldsymbol{\mu_{\tilde{s}}} & =  \boldsymbol{\tilde{A}}\boldsymbol{s}_0 + \boldsymbol{\tilde{B}} \boldsymbol{\tilde{k}} +  \boldsymbol{\tilde{G}} \boldsymbol{\tilde{d}}, \\ \boldsymbol{\tilde{\Sigma}}_{ \boldsymbol{\tilde{s}}} &=  \boldsymbol{\tilde{A}}\boldsymbol{\Sigma}_{\boldsymbol{s}_0} \boldsymbol{\tilde{A}}^{\intercal} + \boldsymbol{\tilde{G}} \boldsymbol{\tilde{\Sigma}}_{ \boldsymbol{\tilde{w}}} \boldsymbol{\tilde{G}}^{\intercal},
\end{align*}
%\begin{align*}
%	 \boldsymbol{\tilde{\Sigma}}_{ \boldsymbol{\tilde{s}}} & =  \boldsymbol{\tilde{A}}\boldsymbol{\Sigma}_{\boldsymbol{s}_0} \boldsymbol{\tilde{A}}^{\intercal} +  \boldsymbol{\tilde{G}} \boldsymbol{\tilde{\Sigma}}_{ \boldsymbol{\tilde{w}}} \boldsymbol{\tilde{G}}^{\intercal},
%\end{align*}
where $\boldsymbol{\tilde{\Sigma}}_{ \boldsymbol{\tilde{w}}}$ are the stacked estimates of the covariance for each time-step, taken under the $N$ samples drawn during the last forward pass. Furthermore, given the feedback gains, we compute the action covariance along the trajectory $$\boldsymbol{\tilde{\Sigma}}_{ \boldsymbol{\tilde{a}}}= \boldsymbol{\tilde{K}} \boldsymbol{\tilde{A}}\boldsymbol{\Sigma}_{\boldsymbol{s}_0} \boldsymbol{\tilde{A}}^{\intercal} \boldsymbol{\tilde{K}}^{\intercal} +  \boldsymbol{\tilde{K}} \boldsymbol{\tilde{G}} \boldsymbol{\tilde{\Sigma}}_{ \boldsymbol{\tilde{w}}} \boldsymbol{\tilde{G}}^{\intercal} \boldsymbol{\tilde{K}}^{\intercal}.$$
%\begin{align*}
%	 \boldsymbol{\tilde{\Sigma}}_{ \boldsymbol{\tilde{a}}}= \boldsymbol{\tilde{K}} \boldsymbol{\tilde{A}}\boldsymbol{\Sigma}_{\boldsymbol{s}_0} \boldsymbol{\tilde{A}}^{\intercal} \boldsymbol{\tilde{K}}^{\intercal} +  \boldsymbol{\tilde{K}} \boldsymbol{\tilde{G}} \boldsymbol{\tilde{\Sigma}}_{ \boldsymbol{\tilde{w}}} \boldsymbol{\tilde{G}}^{\intercal} \boldsymbol{\tilde{K}}^{\intercal}.
%\end{align*}

\subsection{Augmented Objective and Relaxed Chance Constraints}\label{augmendedObj}
We simplify Objective~\eqref{quadrObj} by using the stacked notation and the closed-loop matrices from Section~\ref{sec:closed-loop}
\begin{align*}
	 & J(\boldsymbol{\tilde{s}}, \boldsymbol{\tilde{a}})  = -\mathbb{E} [\boldsymbol{\tilde{s}}^{\intercal} \boldsymbol{\tilde{M}}_{C} \boldsymbol{\tilde{s}}] + \mathbb{E} [2 \boldsymbol{\tilde{s}}_g^{\intercal} \boldsymbol{\tilde{M}} \boldsymbol{\tilde{s}}] - \mathbb{E}
	[\boldsymbol{\tilde{s}}_g^{\intercal} \boldsymbol{\tilde{M}} \boldsymbol{\tilde{s}}_g] ...                                                                                                                                                                                                                \\
	 & ... + \mathbb{E} [2 \boldsymbol{\tilde{s}}_{r}^{\intercal} \boldsymbol{\tilde{K}}^{\intercal} \boldsymbol{\tilde{D}} \boldsymbol{\tilde{K}} \boldsymbol{\tilde{s}}] - \mathbb{E} [2 \boldsymbol{\tilde{a}}_{r}^{\intercal} \boldsymbol{\tilde{D}} \boldsymbol{\tilde{K}} \boldsymbol{\tilde{s}}]
	-\mathbb{E} [2 \boldsymbol{\tilde{k}}^{\intercal} \boldsymbol{\tilde{D}} \boldsymbol{\tilde{K}} \boldsymbol{\tilde{s}}]...                                                                                                                                                                                \\ \nonumber
	 & ...-\mathbb{E}[\boldsymbol{\tilde{s}}_{r}^{\intercal} \boldsymbol{\tilde{K}}^{\intercal} \boldsymbol{\tilde{D}} \boldsymbol{\tilde{K}} \boldsymbol{\tilde{s}}_{r}] + \mathbb{E} [2 \boldsymbol{\tilde{a}}_{r}^{\intercal} \boldsymbol{\tilde{D}} \boldsymbol{\tilde{K}} \boldsymbol{\tilde{s}}_{r}] + 
	\mathbb{E} [2 \boldsymbol{\tilde{k}}^{\intercal} \boldsymbol{\tilde{D}} \boldsymbol{\tilde{K}} \boldsymbol{\tilde{s}}_{r}] ...                                                                                                                                                                            \\                                                                                                                                                                                                                                                   &... - \mathbb{E} [\boldsymbol{\tilde{a}}_{r}^{\intercal} \boldsymbol{\tilde{D}} \boldsymbol{\tilde{a}}_{r}] - \mathbb{E} [2 \boldsymbol{\tilde{k}}^{\intercal} \boldsymbol{\tilde{D}} \boldsymbol{\tilde{a}}_{r}] - \mathbb{E} [\boldsymbol{\tilde{k}}^{\intercal} \boldsymbol{\tilde{D}} \boldsymbol{\tilde{k}}].
\end{align*}
Given that the expectations are of linear-quadratic quantities under Gaussian densities, it is possible to evaluate this objective in closed-form. This objective depends only on the forward terms $\boldsymbol{\tilde{k}}$ and can be reformulated as $\tilde{J}(\boldsymbol{\tilde{k}})$.

Following the relaxation presented in Section~\ref{RelaxingCC} and using the stacked notation we can write the upper and lower state-linear chance constraints as
\begin{align}\label{statesupper}
	\boldsymbol{\tilde{b}}_u - \boldsymbol{\tilde{h}}_u^{\intercal} \boldsymbol{\mu_{\tilde{s}}} -\sqrt{2 \boldsymbol{\tilde{h}}_u^{\intercal}  \boldsymbol{\tilde{\Sigma}}_{ \boldsymbol{\tilde{s}}} \boldsymbol{\tilde{h}}_u}  \odot\boldsymbol{\erf}^{-1}(\boldsymbol{1}-2\boldsymbol{\tilde{\theta}}_u)    & \boldsymbol{\geq}\boldsymbol{0}, \\
	- \boldsymbol{\tilde{b}}_l +  \boldsymbol{\tilde{h}}_l^{\intercal} \boldsymbol{\mu_{\tilde{s}}} + \sqrt{2 \boldsymbol{\tilde{h}}_l^{\intercal}  \boldsymbol{\tilde{\Sigma}}_{ \boldsymbol{\tilde{s}}} \boldsymbol{\tilde{h}}_l} \odot\boldsymbol{\erf}^{-1}(2\boldsymbol{\tilde{\theta}}_l-\boldsymbol{1}) & \boldsymbol{\geq}\boldsymbol{0},
\end{align}
where $\boldsymbol{\tilde{h}}$ and $\boldsymbol{\tilde{b}}$ parameterize the upper and lower half-planes of the state constraints and $ \boldsymbol{\tilde{\theta}}_u $ and $ \boldsymbol{\tilde{\theta}}_l $ denote the probability values per time-step, all stacked and indexed by $u$ and $l$ respectively. 
Analogously, the action constraints of the closed-loop system can be formulated
\begin{align}
	\boldsymbol{\tilde{z}}_{\mathrm {u}} -  \boldsymbol{\tilde{f}}_u^{\intercal} (\boldsymbol{\tilde{K}}(\boldsymbol{\mu_{\tilde{s}}} - \boldsymbol{\tilde{s}}_r)+ \boldsymbol{\tilde{a}}_r +  \boldsymbol{\tilde{k}})- \boldsymbol{\lambda}_u \boldsymbol{\geq}\boldsymbol{0}, \label{actionsupper} \\
	- \boldsymbol{\tilde{z}}_l +  \boldsymbol{\tilde{f}}_l^{\intercal} (\boldsymbol{\tilde{K}}(\boldsymbol{\mu_{\tilde{s}}} -  \boldsymbol{\tilde{s}}_r)+ \boldsymbol{\tilde{a}}_r +  \boldsymbol{\tilde{k}})+ \boldsymbol{\lambda}_l \boldsymbol{\geq}\boldsymbol{0}, \label{actionslower}
\end{align}
where $\boldsymbol{\lambda}_u = \sqrt{2 \boldsymbol{\tilde{f}}_u^{\intercal} \boldsymbol{\tilde{\Sigma}}_{ \boldsymbol{\tilde{a}}} \boldsymbol{\tilde{f}}_u} \odot\boldsymbol{\erf}^{-1}(\boldsymbol{1}-2\boldsymbol{\tilde{\vartheta}}_u) $
and $\boldsymbol{\lambda}_l = \sqrt{2 \boldsymbol{\tilde{f}}_l^{\intercal} \boldsymbol{\tilde{\Sigma}}_{ \boldsymbol{\tilde{a}}} \boldsymbol{\tilde{f}}_l} \odot\boldsymbol{\erf}^{-1}(2\boldsymbol{\tilde{\vartheta}}_l -\boldsymbol{1}) $ , $\boldsymbol{\tilde{f}}$ and $\boldsymbol{\tilde{z}}$ are the stacked half-plane parameters of the action constraints and $ \boldsymbol{\tilde{\vartheta}}_u, \boldsymbol{\tilde{\vartheta}}_l $ are the stacked upper and lower bound probabilities per time-step. The operator $ \odot $ denotes the element-wise multiplication.

\subsection{Chance-Constrained Trajectory Optimization}
Based on the formulations introduced in Section~\ref{sec:closed-loop} and Section~\ref{augmendedObj}, it is possible to construct an optimization problem around the reference trajectory to find a sequence of feedforward terms $\boldsymbol{\tilde{k}}$ that maintain the Constraints~(\ref{statesupper}-\ref{actionslower}).

\begin{figure*}[t!]
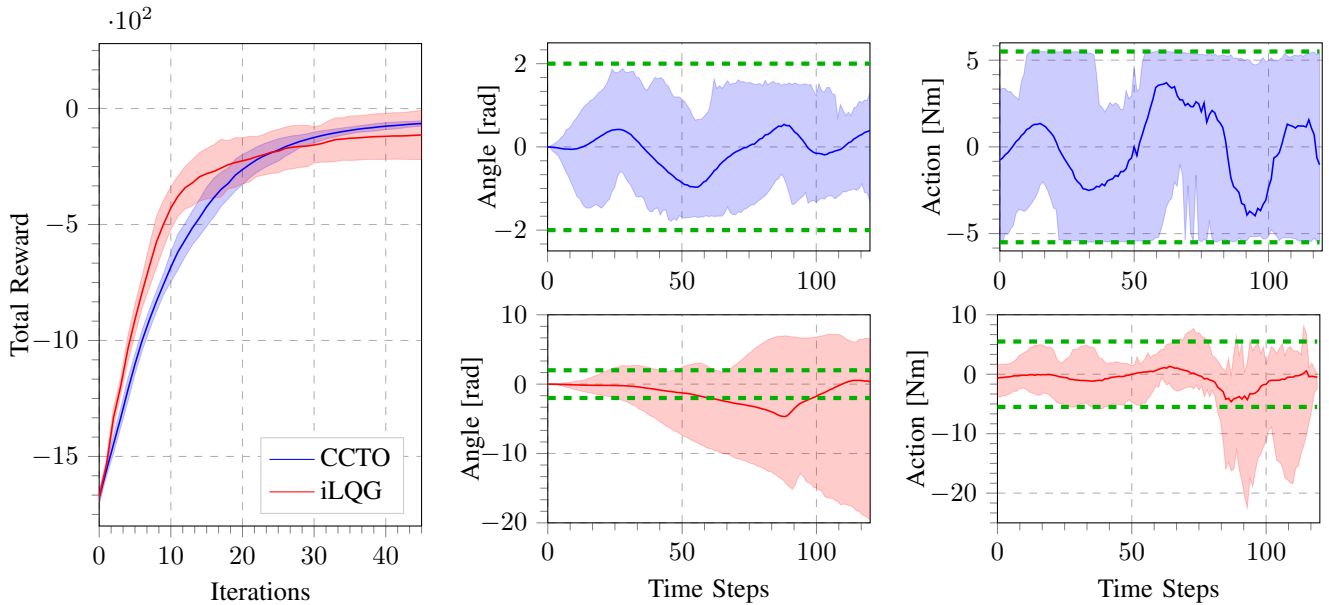

	\centering
	\begin{minipage}[t!]{0.33\textwidth}
		% This file was created by matplotlib2tikz v0.6.18.
\begin{tikzpicture}

\begin{axis}[
legend cell align={left},
legend entries={{CCTO},{iLQG}},
legend style={at={(0.5,0.025)}, anchor=south west, draw=white!80.0!black},
tick align=outside,
tick pos=left,
grid style = dashed,
grid=major,
minor tick num=5,
width=\linewidth,
height=8cm,
try min ticks=5,
scaled y ticks=base 10:-2,
x grid style={white!69.01960784313725!black},
xlabel={Iterations},
xmin=0, xmax=45,
y grid style={white!69.01960784313725!black},
ymin=-1800, ymax=280,
ylabel={Total Reward},
ylabel shift = {-0.2cm}
]
\addlegendimage{no markers, blue}
\addlegendimage{no markers, red}
\path [draw=blue, fill=blue, opacity=0.2] (axis cs:0,-1647.40693375213)
--(axis cs:0,-1705.23192703294)
--(axis cs:1,-1588.71285196325)
--(axis cs:2,-1488.23147044177)
--(axis cs:3,-1375.76879992699)
--(axis cs:4,-1264.8463142994)
--(axis cs:5,-1145.47776639927)
--(axis cs:6,-1044.70430978812)
--(axis cs:7,-956.811140415757)
--(axis cs:8,-875.541426957728)
--(axis cs:9,-802.490440863244)
--(axis cs:10,-747.79940489133)
--(axis cs:11,-695.272043619243)
--(axis cs:12,-639.009023280554)
--(axis cs:13,-588.459174801682)
--(axis cs:14,-550.218886177598)
--(axis cs:15,-507.978758038223)
--(axis cs:16,-468.623338989061)
--(axis cs:17,-423.019418948759)
--(axis cs:18,-390.043972299075)
--(axis cs:19,-355.975192988862)
--(axis cs:20,-328.545663722049)
--(axis cs:21,-299.137061761868)
--(axis cs:22,-275.88011804973)
--(axis cs:23,-256.616426792627)
--(axis cs:24,-240.657890701999)
--(axis cs:25,-227.580632829541)
--(axis cs:26,-205.063257707048)
--(axis cs:27,-188.655993898132)
--(axis cs:28,-169.950041055438)
--(axis cs:29,-157.517356763954)
--(axis cs:30,-148.373811460011)
--(axis cs:31,-138.467681219879)
--(axis cs:32,-128.693923645537)
--(axis cs:33,-121.565113668636)
--(axis cs:34,-114.856244779511)
--(axis cs:35,-108.979875800649)
--(axis cs:36,-103.062308218262)
--(axis cs:37,-100.520094671071)
--(axis cs:38,-98.3618500838873)
--(axis cs:39,-94.8695126746822)
--(axis cs:40,-91.1454685461852)
--(axis cs:41,-88.6876077115191)
--(axis cs:42,-85.6473936000638)
--(axis cs:43,-81.3185718142985)
--(axis cs:44,-77.9215582037346)
--(axis cs:45,-77.0407908719066)
--(axis cs:45,-52.5617132591317)
--(axis cs:45,-52.5617132591317)
--(axis cs:44,-54.8638194580065)
--(axis cs:43,-55.712490964373)
--(axis cs:42,-57.1101822653789)
--(axis cs:41,-58.9238832987185)
--(axis cs:40,-60.9608251531209)
--(axis cs:39,-63.2925682094297)
--(axis cs:38,-67.3763434755832)
--(axis cs:37,-70.9469256254777)
--(axis cs:36,-75.9417351442755)
--(axis cs:35,-79.6711406871712)
--(axis cs:34,-83.4907544317605)
--(axis cs:33,-87.9330952252616)
--(axis cs:32,-92.2205703183863)
--(axis cs:31,-95.9759998788608)
--(axis cs:30,-101.025170116338)
--(axis cs:29,-108.865378912765)
--(axis cs:28,-114.221477748959)
--(axis cs:27,-117.707531270679)
--(axis cs:26,-123.608094816216)
--(axis cs:25,-130.19228766616)
--(axis cs:24,-141.06188115422)
--(axis cs:23,-151.929871383282)
--(axis cs:22,-165.3306493232)
--(axis cs:21,-182.830985284549)
--(axis cs:20,-197.912723941681)
--(axis cs:19,-218.570468176142)
--(axis cs:18,-256.127493333859)
--(axis cs:17,-275.055537096391)
--(axis cs:16,-300.183388436927)
--(axis cs:15,-339.147972152675)
--(axis cs:14,-392.330583934782)
--(axis cs:13,-441.772254595627)
--(axis cs:12,-493.752155739142)
--(axis cs:11,-540.318403110698)
--(axis cs:10,-619.462540618798)
--(axis cs:9,-710.506003066417)
--(axis cs:8,-786.438330146553)
--(axis cs:7,-866.908410227033)
--(axis cs:6,-954.778734086392)
--(axis cs:5,-1058.22078997058)
--(axis cs:4,-1164.40975279119)
--(axis cs:3,-1292.57344890579)
--(axis cs:2,-1407.07548427654)
--(axis cs:1,-1534.61380449299)
--(axis cs:0,-1647.40693375213)
--cycle;

\path [draw=red, fill=red, opacity=0.2] (axis cs:0,-1647.40693375213)
--(axis cs:0,-1705.23192703294)
--(axis cs:1,-1563.4616155871)
--(axis cs:2,-1374.41460191866)
--(axis cs:3,-1233.21992140159)
--(axis cs:4,-1094.44795559316)
--(axis cs:5,-975.59745953762)
--(axis cs:6,-859.871119925035)
--(axis cs:7,-773.238568803988)
--(axis cs:8,-685.384607195163)
--(axis cs:9,-600.331231311893)
--(axis cs:10,-520.465527101869)
--(axis cs:11,-465.45809636701)
--(axis cs:12,-432.632958633898)
--(axis cs:13,-419.905889181029)
--(axis cs:14,-398.30852132218)
--(axis cs:15,-389.961014526558)
--(axis cs:16,-385.693505831241)
--(axis cs:17,-381.454540882159)
--(axis cs:18,-345.54084036019)
--(axis cs:19,-339.77655747592)
--(axis cs:20,-324.745637028972)
--(axis cs:21,-316.601601672865)
--(axis cs:22,-293.613398703536)
--(axis cs:23,-286.403091923345)
--(axis cs:24,-280.192178860552)
--(axis cs:25,-277.351218740204)
--(axis cs:26,-263.231147922537)
--(axis cs:27,-259.612524768675)
--(axis cs:28,-253.902540487832)
--(axis cs:29,-251.532811289983)
--(axis cs:30,-234.346048161264)
--(axis cs:31,-231.423549356102)
--(axis cs:32,-231.07626614609)
--(axis cs:33,-228.132369208172)
--(axis cs:34,-222.980306261468)
--(axis cs:35,-223.720888652115)
--(axis cs:36,-222.515591859706)
--(axis cs:37,-219.95429487084)
--(axis cs:38,-219.148561655312)
--(axis cs:39,-219.590625059059)
--(axis cs:40,-218.950041677301)
--(axis cs:41,-218.463680346093)
--(axis cs:42,-218.463680346093)
--(axis cs:43,-219.090924671036)
--(axis cs:44,-219.728861087721)
--(axis cs:45,-220.139935643944)
--(axis cs:45,-7.87156361478144)
--(axis cs:45,-7.87156361478144)
--(axis cs:44,-11.3023774080153)
--(axis cs:43,-15.2438825175398)
--(axis cs:42,-18.064033756075)
--(axis cs:41,-18.064033756075)
--(axis cs:40,-20.7389653656395)
--(axis cs:39,-21.7503994880707)
--(axis cs:38,-24.6474943089196)
--(axis cs:37,-28.8527931947429)
--(axis cs:36,-31.1300359382368)
--(axis cs:35,-32.3059936072973)
--(axis cs:34,-39.7417902976577)
--(axis cs:33,-40.2494969781532)
--(axis cs:32,-53.1577605452856)
--(axis cs:31,-71.4551288603232)
--(axis cs:30,-79.4481307287798)
--(axis cs:29,-71.9255061642117)
--(axis cs:28,-72.9110346385145)
--(axis cs:27,-73.5790883868232)
--(axis cs:26,-79.98176773976)
--(axis cs:25,-90.2754026679186)
--(axis cs:24,-97.2606679246955)
--(axis cs:23,-111.552587277406)
--(axis cs:22,-121.127248202809)
--(axis cs:21,-119.565238528215)
--(axis cs:20,-126.025873217266)
--(axis cs:19,-125.552623585074)
--(axis cs:18,-139.055265414785)
--(axis cs:17,-132.189874583889)
--(axis cs:16,-156.353586142405)
--(axis cs:15,-171.889664060627)
--(axis cs:14,-190.106875922248)
--(axis cs:13,-221.329174318415)
--(axis cs:12,-249.49663499026)
--(axis cs:11,-290.950470478651)
--(axis cs:10,-336.929765966094)
--(axis cs:9,-394.546442948762)
--(axis cs:8,-464.522286536211)
--(axis cs:7,-589.224498409997)
--(axis cs:6,-722.387110410344)
--(axis cs:5,-842.23751257101)
--(axis cs:4,-981.203017744928)
--(axis cs:3,-1162.58289501416)
--(axis cs:2,-1290.55703229642)
--(axis cs:1,-1513.69223403203)
--(axis cs:0,-1647.40693375213)
--cycle;

\addplot [semithick, blue]
table [row sep=\\]{%
0	-1676.31943039254 \\
1	-1561.66332822812 \\
2	-1447.65347735916 \\
3	-1334.17112441639 \\
4	-1214.62803354529 \\
5	-1101.84927818492 \\
6	-999.741521937255 \\
7	-911.859775321395 \\
8	-830.989878552141 \\
9	-756.49822196483 \\
10	-683.630972755064 \\
11	-617.795223364971 \\
12	-566.380589509848 \\
13	-515.115714698655 \\
14	-471.27473505619 \\
15	-423.563365095449 \\
16	-384.403363712994 \\
17	-349.037478022575 \\
18	-323.085732816467 \\
19	-287.272830582502 \\
20	-263.229193831865 \\
21	-240.984023523208 \\
22	-220.605383686465 \\
23	-204.273149087954 \\
24	-190.85988592811 \\
25	-178.88646024785 \\
26	-164.335676261632 \\
27	-153.181762584405 \\
28	-142.085759402198 \\
29	-133.19136783836 \\
30	-124.699490788175 \\
31	-117.22184054937 \\
32	-110.457246981962 \\
33	-104.749104446949 \\
34	-99.173499605636 \\
35	-94.3255082439102 \\
36	-89.5020216812689 \\
37	-85.7335101482744 \\
38	-82.8690967797352 \\
39	-79.081040442056 \\
40	-76.053146849653 \\
41	-73.8057455051188 \\
42	-71.3787879327213 \\
43	-68.5155313893357 \\
44	-66.3926888308705 \\
45	-64.8012520655191 \\
};
\addplot [semithick, red]
table [row sep=\\]{%
0	-1676.31943039254 \\
1	-1538.57692480956 \\
2	-1332.48581710754 \\
3	-1197.90140820787 \\
4	-1037.82548666904 \\
5	-908.917486054315 \\
6	-791.129115167689 \\
7	-681.231533606993 \\
8	-574.953446865687 \\
9	-497.438837130327 \\
10	-428.697646533981 \\
11	-378.20428342283 \\
12	-341.064796812079 \\
13	-320.617531749722 \\
14	-294.207698622214 \\
15	-280.925339293593 \\
16	-271.023545986823 \\
17	-256.822207733024 \\
18	-242.298052887488 \\
19	-232.664590530497 \\
20	-225.385755123119 \\
21	-218.08342010054 \\
22	-207.370323453173 \\
23	-198.977839600376 \\
24	-188.726423392624 \\
25	-183.813310704061 \\
26	-171.606457831149 \\
27	-166.595806577749 \\
28	-163.406787563173 \\
29	-161.729158727097 \\
30	-156.897089445022 \\
31	-151.439339108213 \\
32	-142.117013345688 \\
33	-134.190933093163 \\
34	-131.361048279563 \\
35	-128.013441129706 \\
36	-126.822813898971 \\
37	-124.403544032791 \\
38	-121.898027982116 \\
39	-120.670512273565 \\
40	-119.84450352147 \\
41	-118.263857051084 \\
42	-118.263857051084 \\
43	-117.167403594288 \\
44	-115.515619247868 \\
45	-114.005749629363 \\
};
\end{axis}

\end{tikzpicture}
		\label{furuta_reward}
	\end{minipage}
	\begin{minipage}[t!]{0.33\textwidth}
		\vspace{5mm}
		\begin{minipage}[]{\linewidth}
			\hspace{1mm}
			\input{graphs/furuta/lin_trajs_furuta_008_CC}
			\label{linear_trajs_CC}
		\end{minipage}\hfill
		\begin{minipage}[]{\linewidth}
			\input{graphs/furuta/lin_trajs_furuta_008_iLQG}
			\label{linear_trajs_iLQG}
		\end{minipage}
	\end{minipage}\hfill
	\begin{minipage}[t!]{0.33\textwidth}
		\vspace{5mm}
		\begin{minipage}[]{\linewidth}
			\hspace{1mm}
			\input{graphs/furuta/actions_furuta_008_CC}
			\label{lin_actions_space_furuta_CC}
		\end{minipage}\hfill
		\begin{minipage}[]{\linewidth}
			\input{graphs/furuta/actions_furuta_008_iLQG}
			\label{lin_actions_space_furuta_iLQG}
		\end{minipage}
	\end{minipage}\hfill
	\setlength{\belowcaptionskip}{-4mm}
	\vspace{-1mm}
	\caption{Total-reward curve reflecting the performance of iLQG and CCTO for the Furuta pendulum swing-up task (left). In addition, we show the space (min. and max.) of planned nominal trajectories of the constrained angle (middle) and the corresponding executed actions (right), CCTO (blue), iLQG (red). CCTO obeys the physical limits of the system, while iLQG drives the dynamics against the constraints (green). These violations lead to poor linear approximations of the dynamics and an overall slightly lower mean and higher variance performance of iLQG.}
	\label{furuta}
\end{figure*}
The resulting optimization is a quadratic program with linear constraints in $ \boldsymbol{\tilde{k}}$. Thus, the probabilistic problem reduces to a deterministic one, which can be solved efficiently with many numerical solvers, for example, qpOASES \cite{ferreau2014qpoases} within the CasADi framework \cite{andersson2018casadi}. The complete dynamic programming and optimization loop is described in Algorithm~1 and is summarized as follows: During an initial forward pass, we obtain $N$ trajectory samples, around which the dynamics is linearized for each time-step. The linearized dynamics is used to perform the backward pass of iLQG and obtain the feedback and feedforward controllers along the reference trajectory. These controllers are then used to formulate the closed-loop linearized system with the stacked notation and to warm-start the quadratic program. The solution of the constrained program returns the optimal feedforward sequence $\boldsymbol{k}_t$, which is used to perform the next forward pass and linearization. Following \cite{tassa2012synthesis}, we also use the hyperparameter $\alpha$ that scales the feedforward control in order to keep the next forward pass of the non-linear system in a valid trust-region around the linear-quadratic approximations.

\begin{algorithm}[t]\label{algo}
	\caption{Chance-Constrained Trajectory Opt. (CCTO)}
	\begin{algorithmic}[1]
		\renewcommand{\algorithmicrequire}{\textbf{Input:}}
		\renewcommand{\algorithmicensure}{\textbf{Output:}}
		\REQUIRE $\theta_{u, t},~\theta_{l,t},~\vartheta_{u,t},~\vartheta_{l,t}, ~\alpha , N$
		\ENSURE  $\boldsymbol{K}_t,~ \boldsymbol{k}_t,~ \boldsymbol{s}_{r,t},~ \boldsymbol{a}_{r,t} $
		\STATE $\boldsymbol{a}_{t}^{1:N}, \boldsymbol{s}_{t}^{1:N} \leftarrow $ forwardPass($\boldsymbol{a}_{r,t}$, $\boldsymbol{s}_{r,t}$, $\boldsymbol{K}_t $, $\boldsymbol{k}_t $, $\alpha $)
		\WHILE{not converged}
		\STATE $\boldsymbol{a}_{r, t}, \boldsymbol{s}_{r, t} \leftarrow $ meanTraj($\boldsymbol{a}_{t}^{1:N}$, $\boldsymbol{s}_{t}^{1:N}$)
		\STATE $\boldsymbol{A}_t $, $\boldsymbol{B}_t$, $\boldsymbol{c}_t  \leftarrow$ fitDynamics($\boldsymbol{a}_{t}^{1:N}, \boldsymbol{s}_{t}^{1:N}$)
		\STATE $\boldsymbol{K}_t,~ \boldsymbol{k}_t^{\star} \leftarrow $ backwardPass($\boldsymbol{A}_t $, $\boldsymbol{B}_t $)
		\STATE $\boldsymbol{k}_t$ \hspace{-2mm} $\leftarrow$ \hspace{-2mm} solveQP($\boldsymbol{A}_t, \boldsymbol{B}_t, \boldsymbol{c}_t, \boldsymbol{K}_t, \boldsymbol{k}_t^{\star}, \theta_{u,t},\theta_{l,t},\vartheta_{u,t},\vartheta_{l,t}$)
		\STATE $\boldsymbol{a}_{t}^{1:N}, \boldsymbol{s}_{t}^{1:N} \leftarrow $ forwardPass($\boldsymbol{a}_{r,t}$, $\boldsymbol{s}_{r,t}$, $\boldsymbol{K}_t $, $\boldsymbol{k}_t $, $\alpha $)
		\ENDWHILE
	\end{algorithmic}
\end{algorithm}
\setlength{\textfloatsep}{0pt}% Remove \textfloatsep

\vspace{-4mm}
\section{Empirical Evaluation}
We evaluate our approach on two highly non-linear dynamical tasks, the Furuta pendulum \cite{furuta1992swing} and a Cart-Pole environment. Both problems are under-actuated and have state and actions constraints. We consider quadratic reward functions for both experiments and set the probability values for violating the constraints to $\theta_u = \theta_l =\vartheta_u =\vartheta_l = 0.01 $.

\begin{figure*}[t!]
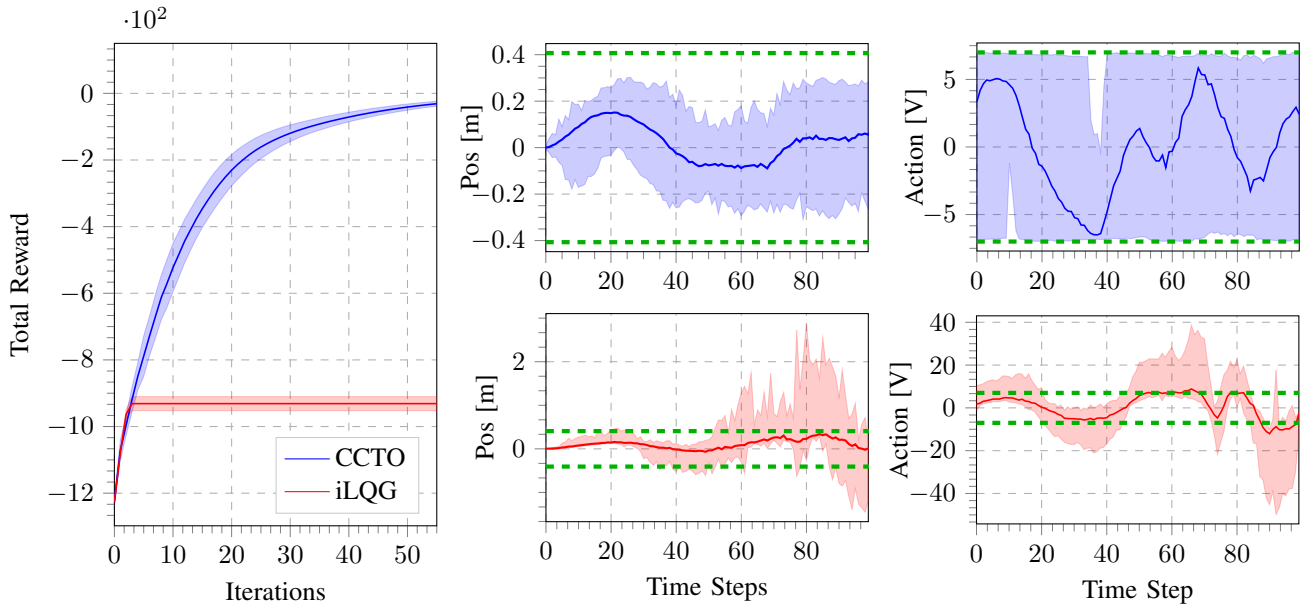

	\centering
	\begin{minipage}[t!]{0.33\textwidth}
		\input{graphs/cartpole/rewards_cartpole_01}
		\label{reward_cartpole}
	\end{minipage}\hfill
	\begin{minipage}[t!]{0.33\textwidth}
		\vspace{0mm}
		\begin{minipage}[]{\linewidth}
			\hspace{0mm}
			\input{graphs/cartpole/lin_trajs_cartpole_CC}
			\label{linear_trajs_CC_cartpole}
		\end{minipage}
		\begin{minipage}[]{\linewidth}
			\vspace{2mm}
			\hspace{2mm}
			\input{graphs/cartpole/lin_trajs_cartpole_iLQG}
			\label{linear_trajs_iLQG_cartpole}
		\end{minipage}
	\end{minipage}\hfill
	\begin{minipage}[t!]{0.33\textwidth}
		\vspace{2mm}
		\begin{minipage}[]{\linewidth}
			\vspace{-1mm}
			\hspace{0mm}
			\input{graphs/cartpole/actions_cartpole_01_CC}
			\label{lin_actions_space_cartpole_CC}
		\end{minipage}\hfill
		\begin{minipage}[]{\linewidth}
			\vspace{1mm}
			\hspace{-2mm}
			\input{graphs/cartpole/actions_cartpole_01_iLQG}
			\label{lin_actions_space_cartpole_iLQG}
		\end{minipage}
	\end{minipage}\hfill
	\setlength{\belowcaptionskip}{-4mm}
	\vspace{-5mm}
	\caption{Total-reward curve reflecting the performance of iLQG and CCTO for the Cart-Pole swing-up task (left). Furthermore, we show the space (min. and max.) of planned nominal trajectories of the constrained position (middle) and the corresponding executed actions (right), CCTO (blue), iLQG (red). CCTO obeys the physical limits of the system, while iLQG drives the dynamics against the constraints (green). These violations, especially those of the action constraint cause iLQG to get stuck in a poor local optimum, while CCTO is able to solve the task and perform the swing-up.}
	\label{cartpole}
\end{figure*}

\paragraph{Furuta Pendulum Swing-Up}In the Furuta pendulum the state is represented by the angles of both links and the corresponding angular velocities. Only the horizontal link is actuated and is subject to both state and the action constraints. To make the environment stochastic, we introduce both action and process noise. We run our experiment under identical conditions for CCTO and iLQG. We fix the feedforward scalar $\alpha$ to $0.05$ for both algorithms and perform 20 seeded trials, each with 45 iterations, 50 rollouts per iteration. The resulting performance curve of both algorithms can be seen in Figure~\ref{furuta}. Furthermore, we present the planned nominal trajectories, as well as the planned nominal actions of both algorithms for one trial. The filled space is the area between the minimum and maximum values of states and actions and should not be confused with a probability distribution over trajectories. The advantage of our approach is clear. CCTO reaches better overall performance with a higher final reward and smaller standard deviation, Table~\ref{mean_reward_std_furuta}. iLQG plans frequently and consistently to violate the constraints, while CCTO keeps the state and action trajectories within a feasible space. This consideration leads to an improved approximation of the non-linear system dynamic and allows CCTO to perform robust improvement steps during the optimization. This result is affirmed by the low regularization values of CCTO, Table~\ref{Reg_params_table}.
%\begin{figure}[h!]
%	\centering
%	\includegraphics[width=0.7\columnwidth]{graphs/cartpole/cartpole}
%	\caption{The cartpole during a swing-up with CCTO.}
%	\label{cartpole}
%\end{figure}
\paragraph{Cart-Pole Swing-Up}
For the well-known Cart-Pole environment, we consider constraints on the position of the cart as well as on the action. To make the task more challenging, we again apply action and process noise, enforce harsh action constraints and limit the time horizon to 100 time steps, the equivalent of 2 seconds. We evaluate iLQG and CCTO on 20 seeded trials, each with 55 iterations and 50 rollouts per iteration. We set the feedforward scaling parameter $\alpha$ to $0.1$. Analogously to the last experiment, Figure~\ref{cartpole} depicts the performance curve of iLQG and CCTO, as well as the spaces of planned nominal trajectories for the cart's position and the corresponding actions. In this experiment, iLQG moves very quickly towards a local optimum and does not manage to swing the Cart-Pole up. In contrast, CCTO performs the swing-up by finding a suitable nominal trajectory in the feasible constrained space. Tables~\ref{mean_reward_std_cartpole} and \ref{Reg_params_table_cartpole} reflect the performance discrepancy between both algorithms, in terms of total rewards and needed regularization.

\begin{table}[t]
	\begin{minipage}[t]{0.45\textwidth}
		\centering
		\def\arraystretch{1.25}% 
		\begin{tabular}{| l | c | c | c | c |}
			\hline
			Iteration & 10                        & 30                        & 45                        \\ \hline
			CCTO      & $-6.8 (\pm0.32)$          & $\mathbf{-1.3(\pm0.11)} $ & $\mathbf{-0.65(\pm0.6)} $ \\ \hline
			iLQG      & $\mathbf{-4.3(\pm0.46)} $ & $-1.6(\pm0.39) $          & $-1.1(\pm0.53) $          \\ \hline
		\end{tabular}
		\caption{Mean total reward and standard deviation of the Furuta swing-up task scaled by $1\mathrm{e}{-2}$.}
		\label{mean_reward_std_furuta}
	\end{minipage}\hfill
	\vspace{2mm}
	\begin{minipage}[t]{0.95\columnwidth}
		\centering
		\def\arraystretch{1.25}% 
		\begin{tabular}{ | l | c | c | c | c |}
			\hline
			Iteration & 10            & 30                            & 45                          \\ \hline
			CCTO      & $\mathbf{0} $ & $\mathbf{2.5\mathrm{e}{-8}} $ & $\mathbf{1\mathrm{e}{-4}} $ \\ \hline
			iLQG      & $\mathbf{0} $ & $ 2.85\mathrm{e}{38} $        & $ 5\mathrm{e}{80} $         \\ \hline
		\end{tabular}
		\caption{Mean regularization in the Furuta task over all trials for different iterations.
			CCTO needs less regularization due to avoidance of hard non-linearities.}
		\label{Reg_params_table}
	\end{minipage}%
	\vspace{2mm}
\end{table}

\begin{table}[t]
	\begin{minipage}[t]{0.45\textwidth}
		\centering
		\def\arraystretch{1.25}% 
		\begin{tabular}{| l | c | c | c |}
			\hline
			Iteration & 20                       & 30                        & 55                         \\ \hline
			CCTO      & $\mathbf{-2.3(\pm0.32)}$ & $\mathbf{-1.2(\pm0.32)} $ & $\mathbf{-0.31(\pm0.06)} $ \\ \hline
			iLQG      & $-9.3(\pm0.10) $         & $-9.3(\pm0.10) $          & $-9.3(\pm0.10) $           \\ \hline
		\end{tabular}
		\caption[]{Mean total reward and standard deviation of the Cart-Pole swing-up task scaled by $1\mathrm{e}{-2}$.}
		\label{mean_reward_std_cartpole}
	\end{minipage}\hfill
	\vspace{2mm}
	\begin{minipage}[t]{0.45\textwidth}
		\centering
		\def\arraystretch{1.25}% 
		\begin{tabular}{| l | c | c | c | c |}
			\hline
			Iteration & 20                   & 30                  & 55                  \\ \hline
			CCTO      & $\mathbf{0} $        & $\mathbf{0} $       & $\mathbf{0} $       \\ \hline
			iLQG      & $5.7\mathrm{e}{39} $ & $ 1\mathrm{e}{80} $ & $ 1\mathrm{e}{80} $ \\ \hline
		\end{tabular}
		\caption[]{Mean regularization in the Cart-Pole task over all trials for different iterations.
			CCTO needs less regularization due to avoidance of hard non-linearities.}
		\label{Reg_params_table_cartpole}
	\end{minipage}%
	\vspace{2mm}
\end{table}

\section{Conclusion and Future Research}
We have proposed a new trajectory optimization technique, based on the framework of differential dynamic programming, that takes into consideration probabilistic chance constraints in stochastic environments with unknown non-linear dynamics. We used Boole's inequality to conservatively relax the non-convex chance constraints, enabling us to formulate a constrained quadratic program and optimize the nominal trajectory to stay inside the feasible set defined by the probabilistic linear state and action limits. We have provided a thorough derivation of our approach and empirically demonstrated the advantage of enforcing physical limits on two simulated highly dynamical and stochastic non-linear systems. The results indicate that incorporating the chance constraints leads to higher fidelity in the online-fitted local linear-quadratic approximations, and consequently greatly influences the robustness of the iterative optimization process. This observation is reflected in very low regularizations in comparison to standard iLQG.

In future research, we will extend our optimization to include not only the nominal trajectory but also the feedback gains, and we will consider optimizing the probabilistic constraint bounds via risk allocation to achieve dynamic risk measures across time and iterations. In addition, we plan to move to the fully stochastic optimization framework of maximum-entropy iLQG \cite{levine2014learning} to avoid regularization heuristics of the DDP framework.

\vspace{-1.5mm}
\bibliographystyle{IEEEtran}
\bibliography{references}

\end{document}